# Psychophysical identity and free energy

Alex B. Kiefer
Monash University

**ABSTRACT**

An approach to implementing variational Bayesian inference in biological systems is considered, under which the thermodynamic free energy of a system directly encodes its variational free energy. In the case of the brain, this assumption places constraints on the neuronal encoding of generative and recognition densities, in particular requiring a stochastic population code. The resulting relationship between thermodynamic and variational free energies is prefigured in mind-brain identity theses in philosophy and in the Gestalt hypothesis of psychophysical isomorphism.

**Introduction**

In machine learning and, increasingly, in cognitive neuroscience, it is widely recognized that the bulk of the learning undergone during an organism's lifetime is likely to be unsupervised—that is, to be based directly on the modeling of streams of incoming sensory data, rather than on explicit reinforcement of downstream performance. Theories of unsupervised learning in the brain invariably appeal, in one way or another, to the iterative refinement of a neuronally implemented, hierarchically organized generative model of the sensory data. The cost function used to update this model, which measures the difference between actual and predicted sensory inputs, is often formally analogous, in some cases quite precisely, to descriptions of physical systems in terms of their potential energy [1]. Variational methods, which formalize statistical inference and learning in terms of the maximization of a lower bound on the model evidence called (negative) variational free energy, are among state-of-the-art approaches in this vein [2], and have a long history in the theory of unsupervised learning in artificial neural networks [3,4].

In theoretical neuroscience, variational free energy (VFE) minimization has been proposed as a unifying explanatory framework accounting in principle for all psychologically significant aspects of cortical function, particularly those underwriting perception and action [5-8]. This theoretical approach, sometimes called the "free energy principle" (FEP), has recently been extended from a theory of the brain to a more general emerging framework that treats life in general, at all spatiotemporal scales and developmental stages, in terms of a gradient descent on free energy [9-13].[1]

In this paper, I consider a way of implementing variational inference in the brain that is non-accidentally related to its thermodynamic description. The implications of this perspective on variational inference, I argue, are similar to those of positions taken in 20th-century philosophy on the mind-brain relation [14,15]. J.J.C. Smart's [14] "topic-neutral" analysis of mental state ascriptions, for example, allowed for the possibility that mental states are (despite initial appearances, perhaps) brain states. Similarly, on the present view, the mathematics of variational inference delineates a set of formal relations that obtain within a system whether it is described cognitively, in terms of the contents of its computational states, or purely physically, in terms of the energy dynamics governing the (neuronal) vehicles of its representations.

This theoretical stance is stronger than that adopted historically by proponents of the FEP, who have in some cases underscored the model-relative status of the free energy of approximate Bayesian inference [16,17] and its distinctness from thermodynamic free energy. Recent work on this topic by Karl Friston and colleagues, however, confirms a systematic link between the VFE and thermodynamic potential energy. In [18] it is demonstrated that changes in thermodynamic potential energy are approximately equal to changes in the surprisal or self-information of a system near its non-equilibrium steady states, and [9] (see pp.65-67) describes the minimization of thermodynamic free energy as a system approaches non-equilibrium steady state. [19] draws a conclusion similar to that of the present work via a somewhat different route.[2]

Interestingly, this view was presaged not only by the identity theorists in philosophy but also, quite precisely, by the Gestalt psychologists, who supposed that perceptual phenomena as subjectively experienced had structures isomorphic to their underlying physiological correlates (see e.g. [20], p.56, [21], p.552, and [22]).[3]

The outline of this paper is as follows. In §1, I introduce the variational approach to Bayesian inference, briefly discuss its technical motivation, and describe its formal and conceptual relations to the notion of free energy from statistical mechanics. In §2, I consider how probabilities (and state updates) would need to be encoded in neuronal dynamics such that the same mathematical description can be applied to the former and to the latter, a necessary condition on the truth of the identity thesis. In §3 I discuss how this thesis, which is fundamentally a philosophical claim, may be grounded in the mathematics via Lewisian functionalism. I conclude by considering the scope and implications of the thesis.

---

[1] These twin developments are of course intimately related, as modeling in computational neuroscience has come increasingly to be informed by progress in those branches of machine learning that themselves take ongoing inspiration from neuroscience (i.e. deep learning / connectionism).

[2] The thesis presented in this paper was developed independently of Friston et al's work in [19], though the two approaches are clearly complementary.

[3] The Gestalt theorists argued for this isomorphism from the holistic ("molar") nature of both perceptual experience and physical processes, supposing the brain to behave more or less as a unified electrostatic field in which modification of one part necessarily modifies all the others. Later in his career, Köhler shifted to a version of the view rooted in synaptic potentials, which seems not to be falsified by the experiments historically used to reject the isomorphism thesis [23]. In any case, the present considerations amount to an alternative argument for a similar conclusion to the Gestaltists'.

**1 The formal equivalence**

*1.1 Free energy in optimization and in thermodynamics*

Free energy, in the context of variational Bayesian inference, is a function(al) of a probability distribution or density $Q$ used to approximate the (in practice, typically intractable) joint posterior distribution $P(H, V)$ of data $V$, together with the (unobserved) cause(s) of those data, $H$, under a statistical model whose parameters may also be unknown.[4] In relevant applications in cognitive science, variational inference has been proposed as a mechanism whereby the inputs to a neuronal processing hierarchy, originating in sensory transducers, are used to infer a set of neuronally encoded hypotheses about the causes of the sensory input stream, given knowledge only of the sensory inputs and an evolving empirical prior over their causes. In this context, the approximating distribution[5] is typically described as a "recognition" model or density (since it can be used to recognize causes given sensory inputs) and the approximated distribution as a "generative model" (since it can be used to generate states of or predictions for the system's input channel(s), a process that mirrors the causation of sensory input by external sources).

The generative model, which in theories such as hierarchical predictive coding [25] is hypothesized to be implemented in top-down cortical connections, specifies the *Umwelt* of the organism, the kinds of things and situations it believes in independently of the current sensory data (in the literature on active inference, states with high probability under the generative model are sometimes called "phenotypic states", since they are the states that a creature of a given kind must be in to remain viable [26]). Neuronal responses to sensory inputs, as well as the organism's ongoing interactions with the environment, have the effect of increasing the goodness of fit between the generative density and the inferred states on average, both by changing the world to fit the generative density over causes and by modifying the generative density itself in response to observations [10].

---

[4] Throughout, I have adopted the convention from connectionist modeling of using '$v$' for an observed or "visible" variable, and '$h$' for a "hidden" or unobserved one.

[5] As noted in [24, p.463], variational inference may be used to find exact solutions. However, by restricting the form of $Q$ (for example to a factorial distribution), one gains tractability (and in the present context, biological plausibility) at the expense of accuracy.

Against this background, free energy captures the discrepancy between the organism's generative model of the world and the current environmental conditions, where the latter are represented most immediately by the approximating recognition density. If we perform inference over the parameters $\theta$ of the generative model as well as the latent variables or "hidden states" $H$, subsuming both under a random variable $Z$, the negative free energy, $F$, can be written in the following form (closely following [24], but with the dependence of each $Q(z)$ on a particular observed variable $v$ made explicit):

$$F = \sum_Z Q(z|v) \log\left(\frac{P(v,z)}{Q(z|v)}\right) \qquad (1)$$

Note that the true joint probability of latent and observed variables $P(v, h)$ is parameterized by $\theta$, so that $F$ is a function of $Q$ and $\theta$—and since the parameters $\theta$ are subsumed under $Z$, implicitly represented here by $Q(z|v)$, $F$ depends only on $Q$.

$F$ is useful as an optimization target for several reasons that have been widely discussed in relevant literature. Key properties are summarized in Eq. (2), where $L = \log P(v)$, the log probability of the data $v$ under a generative model of their causes, and the first term is $D_{KL}(Q(z|v)\|P(z|v))$, the K-L divergence between $Q$ and the posterior distribution over latent variables and generative parameters:

$$\sum_Z Q(z|v) \log\left(\frac{Q(z|v)}{P(z|v)}\right) + F = L \qquad (2)$$

The K-L divergence quantifies the difference between $P$ and $Q$, is zero when they are identical, and is otherwise positive. Thus, $F$ acts as a lower bound on the log likelihood of the data under $P$. Moreover, holding $L$ fixed, maximizing $F$ necessarily minimizes $D_{KL}$. These are two aspects of a single optimization process in which the data distribution alone is used to infer the best possible model[6] of it and its hidden causes, making variational inference suitable as a method for unsupervised learning.

As is often remarked in discussions of variational inference, $F$ has almost precisely the form of a negative (Helmholtz) free energy from statistical mechanics [1,3,4]. This energy may be written as $F(T) = \langle E(T) \rangle - TS$, where the first term on the right is the expectation, across alternative possible states, of the energy stored in the system due to its internal configuration (at temperature $T$), and $TS$ is the temperature times the entropy $S$ ([27], p.673). To see the equivalence, we rewrite Eq. (1) using an "energy" term $E(v, z)$ defined as the negative log joint probability of $v$ and $z$ under the generative model (whose parameters are again treated as stochastic variables). Reversing signs and expanding the log expression gives the following (using the label *VFE* for the (variational) free energy, as opposed to the negative free energy, $F$):

$$VFE = \sum_z Q(z|v)E(v,z) - \left(-\sum_z Q(z|v)log(Q(z|v))\right) \quad (3)$$

---

[6] One criterion for the goodness of a model is its simplicity, which can be quantified in terms of its Shannon description length [1,4]. Crucially, this is equal to the surprisal associated with the latent variables of the model.

Here, the second term is the entropy of the distribution $Q$ and the first term is the expected "energy" under $Q$. This has the same form as $F(T)$ above assuming a temperature of 1 [4].

*1.2 The role of internal energy*

The present interest in VFE within theoretical neuroscience is attributable to at least two historical influences. The first is the application of algorithms for finding low-energy states of matter to optimization problems [27,28], combined with the assumption that the brain implements a multivariate statistical model [29,30]. A second is Hopfield's observation that the probability of an action potential in a neuron is a smooth (sigmoidal) function of the (short-term average) potential across the cell's membrane ([31], p.2555−see Fig. 1), which is a measure of potential energy (per unit charge). Hopfield showed that networks whose dynamics minimize a global energy function, defined additively in terms of local potentials, can exhibit spontaneously emerging self-organizational properties useful for the storage of memories. Later work synthesized Hopfield's approach with Bayesian inference [32], ultimately yielding models like the Helmholtz machine [33], in which online unsupervised learning is based explicitly on the minimization of a variational free energy. More recently, variational autoencoders [2] and a variety of energy-based models [34,35,36] have made use of similar machinery.

Analogies are, almost by definition, partial. If the connection between statistical mechanics and statistical modeling by the brain were merely one of analogy, it would be surprising to find that all the terms in the Helmholtz free energy play useful and interlocking representational roles. This coincidence between physical and representational descriptions is to be expected, however, if the free energy simply measures how much useful "representational work" can be done by the internal elements of a system, where "work" has its physical meaning. In the remainder of §1 I consider how the physical interpretation of each term in the Helmholtz free energy can be related to a corresponding facet of the optimization process.

A standard expression for the Helmholtz free energy is $A = U - TS$. The internal energy $U$ combines all the energy (potential and kinetic) residing in the system. As we have seen, the energy of a system in statistical mechanics is cast as an average energy over possible configurations, weighted by their probability. The Boltzmann distribution (Eq. 4) relates the energy of a state *s* to its probability at a given temperature:

$$P(s) = \frac{e^{-\frac{E(s)}{k_B T}}}{\sum_{s'} e^{-\frac{E(s')}{k_B T}}} \qquad (4)$$

where *s* is a particular state of the system with energy *E(s)*, the *s'* are the other possible configurations of the system, $k_B$ is Boltzmann's constant, and *T* is temperature.[7] Though the contrast is much more pronounced at lower temperatures, this equation ensures that relatively low-energy states are higher in probability.

This inverse relationship between probability and energy is exploited in most uses of "energy" as a cost function in optimization problems. By construction, low energies are associated with "good" or desired configurations of the system, which are often interpretable as assigning high probability to what they represent. In certain types of stochastic network, the analogy to physics is closer still: the energy can be directly related to the probability of an internal state of the network occurring, as in the Helmholtz machine trained using the stochastic wake-sleep algorithm [4,33], where the probability of a given hidden-layer representation *h*, given a data point *v*, is given by the Boltzmann distribution at the free energy minimum.

*1.3 The roles of entropy and temperature*

The *free* energy is defined by Helmholtz as that portion of a system's available energy "convertible without limit into other work-equivalents" ([37], p.43). This excludes the "bound" portion of the energy associated with heat, represented as the product of the system's entropy *S* and its temperature *T*. An intuitive explanation of the *–TS* term is that, insofar as the properties of the particles in a system are uncertain, their kinetic energy constitutes "irregular motion", so the impact on the free energy of entropy, a "measure of the irregularity" ([37], p.56), is scaled by temperature (roughly, average molecular kinetic energy).

Later formulations of entropy in statistical mechanics use a formula identical to Shannon entropy in information theory, apart from the scale introduced by the Boltzmann constant. Jaynes [38] proposes an influential interpretation of statistical mechanics according to which thermodynamic and information-theoretic entropy in fact "appear as the same *concept*" (p.621), and shows that the Boltzmann distribution falls out as a special case of the principle that the least biased distribution compatible with current knowledge is the one with maximum entropy.

---

[7] The denominator $\sum s' \, e^{-E(s')/k_B T}$ is the partition function *Z*. The Helmholtz free energy is equal to $-k_B T \log Z$, which is minimized when the partition function sums to 1. At this point, the probability of a given state is just an exponential function of its (negative, scaled) energy.

Geoffrey Hinton and colleagues [1,4,39] have argued that it is preferable to use the full Helmholtz free energy as a cost function for learning a recognition distribution $Q$ in a stochastic neural network, rather than simply setting $Q$ to maximize the probability of picking the lowest-cost code (i.e. configuration of the hidden units) for an input vector $v$. This is because the entropy of the distribution over codes, which appears as "free choice" from the point of view of an information source, can be leveraged to communicate surplus information (i.e. beyond that needed to encode $v$). If we have two equally efficient codes for $v$ and use them with equal probability, our choice of code communicates one "extra" bit of information for free. This is of course consistent with the more general argument considered above [38].

The meaning of temperature within statistical modeling is illustrated in the example of simulated annealing, where increasing the temperature increases the variance of the distribution over configurations of the system, and lowering it collapses the distribution to a small range of states near the ground state. Controlling the variance of a distribution is useful in many applications. Language models, for example, can produce much more creative and amusing, if not always grammatically correct, samples at high temperatures.[8]

One can of course decrease the Helmholtz free energy simply by cranking up the heat. Doing so would, however, in the limit destroy the traces of whatever forces were exerted on the system in the first place to create the internal (potential) energy usefully exploitable as work. As just discussed, "cooking" a model decreases its precision, diluting it in the direction of a flat distribution. Decreasing precision also increases entropy, but not necessarily in a way that preserves prior knowledge encoded in the model.

*1.4 The role of equilibrium*

Free energy minimization has been used to explain how organisms manage to keep themselves away from thermodynamic equilibrium with respect to the external environment, i.e. how they maintain themselves in non-equilibrium steady state or homeostasis [10]. Descent into thermodynamic equilibrium, for an organism, is death, and entails a state of maximum entropy or disorganization relative to its phenotypically expected states [26]. By contrast, a relatively low-entropy distribution over its states has been used by many researchers as a criterion for a system's being alive [11,40] or "viable" [41]. It may thus seem obvious that finding the "equilibrium" states of a generative model (as in, e.g., [32]) cannot be coextensive with a thermodynamic energy-minimization process.

---

[8] See for example Andrej Karpathy's well known essay on recurrent neural networks at http://karpathy.github.io/2015/05/21/rnn-effectiveness/.

A first step toward addressing this concern is to adopt a certain tripartite taxonomy of states. First, a *steady state* in the relevant sense is one in which ongoing internal processes with cyclical effects upon one another have evolved to a condition in which concentrations (and thus gradients) of relevant resources are stable—that is to say, net flows of forces into and out of the system's substates are in balance. Maintaining steady state requires ongoing addition of energy from external sources to counter the dissipation of energy to the environment.

A *thermodynamic equilibrium state*, such as the hypothetical heat-death of the universe, on the other hand, is only one very unusual type of steady state in which energy is maximally dispersed and activity all but ceases, because every microscopic exchange is perfectly balanced by another ("detailed balance"). A third relevant type of state, which I'll call "excited" or "perturbed", results when one begins with a steady state and adds energy in excess of that required to maintain homeostasis.

After a perturbation, a system will, *ceteris paribus*, follow a trajectory from the perturbed state back to a steady state. On the other hand, steady states will begin to devolve toward equilibrium states if not enough energy is *added* to the system.[9] The crucial point for present purposes is that the descent from a perturbed state to a steady state, and from a steady state to an equilibrium state, share the same rough trajectory: this series of transitions entails the system's exhausting increasingly more of its potential energy, descending a free energy gradient.

Thus, moving toward steady state is, qualitatively speaking, no different from moving toward equilibrium. There is still a potential problem for the identity thesis, however, in that energy minimization algorithms used for optimization often model physical processes in which the energy is taken right down to the lowest-energy or "ground" states of the system (which occur only at the lowest temperatures). Brains and other biological systems operate in high-temperature regimes by comparison.

Fortunately, statistical learning does not require that the VFE reach a global minimum. In absolute terms, free energy can be minimized not only during perception and perceptual learning but on ontogenetic and phylogenetic timescales as well [10,42]. Algorithms simulating only partial reductions in free energy are at the heart of many proven optimization techniques—see for example [4], in which the E-M algorithm is recast as a matter of incremental free energy minimization, so as to justify partial applications of the E and M steps, or contrastive divergence learning [43], in which the equilibrium distribution of a Restricted Boltzmann Machine is replaced by the distribution after just a few steps of Gibbs sampling. In this case, the contrast between the energy induced by an input vector and the energy after a few steps of the Markov chain induces a gradient sufficient for learning.

---

[9] In practice, of course, living systems will never be in perfectly steady states, but rather recovering toward such states from a relative "excess" or "deficiency" of energy.

The target thesis of this paper is the claim that biological systems, whose internal states come to encode statistical models as a result of spontaneous self-organization in response to environmental pressures [44], learn and make use of these models by minimizing their *physical* free energies. The identity thesis does not specify precisely which approach to variational inference is thereby implemented, or the minimum value the VFE must take. It may be, for example, that biological systems run an algorithm closer to contrastive divergence than to simulated annealing.

## 2  A transparent code

*2.1 Stochastic encoding and variational inference*

In order for the thesis of this paper to make sense, it must be kept in view that the generative and recognition densities of variational inference are densities over (possible) *external causes* of the sensory input. That is to say, in the parlance of most philosophers and nearly all cognitive scientists, they are *representations*, and it is their representational function that defines them as statistical models. Fixing the encoding of the recognition and generative densities allows us to directly relate the "representational work" done as the divergence between the densities is decreased to physical work [45].

In the most general formulation of the free energy principle, the generative model is not assumed to be directly encoded in a system's internal states, but rather specifies which of the system's states are expected under its non-equilibrium steady state distribution [9,26,46]. Sophisticated forms of representation and control, however, require "deep" hierarchical models in which explicit markings of statistical regularities at each level play the role of "sensory input" for the next [47]. I will therefore assume in the remainder that a generative model is at least implicitly encoded in a system's internal states.

It is typically taken for granted in discussions of predictive coding in the brain [25] that hierarchical generative (and recognition) models correspond to the hierarchical functional organization and directionality of cortical networks. For example, "backward" or top-down connections are supposed to parameterize the generative model, while the "forward" or bottom-up connections mediate recognition or fast approximate posterior inference [4,5].

Friston's influential exposition of predictive coding [5,48] assumes a deterministic representation in which densities are encoded by variables representing their sufficient statistics. This approach has the important consequence that, as noted in ([9], p.118), the variational free energy is a property of the current *state* of a system—while the thermodynamic free energy is defined in terms of an expectation over an *ensemble* (or set of possible alternative states). This would seem to preclude a strict identity claim. A further reason to doubt that the VFE can be directly encoded in the Helmholtz free energy of a system is that the latter involves a single distribution over states, whereas the VFE is characterized in terms of a recognition density that approximates an underlying "true" generative posterior.

An identity thesis could be defended, however, if the VFE is encoded stochastically in a single thermodynamic free energy functional that combines the influence of the generative and recognition densities. Many early connectionist models, including the Bayesian Hopfield-style network analyzed in [32], the Helmholtz machine, and the Restricted Boltzmann Machine [49], lend themselves to a simple and natural encoding scheme under which the objective probability of a unit's being in a given state represents the *subjective* (Bayesian, model-relative) probability assigned to the event represented by that state. This representation could provide the basis for a "transparent" encoding of variational free energy.

To consider a system whose analysis is tractable, we may begin with the stochastic wake-sleep algorithm, in which the activities of a single set of units representing the hidden causes of observations (a "total representation" in the terms of [4]) are driven bottom-up by the recognition weights during a "wake" phase of the algorithm and top-down by the generative weights during the "sleep" phase. If "wake" and "sleep" phases occur with probabilities $S(\text{wake})$ and $S(\text{sleep}) = 1 - S(\text{wake})$ respectively,[10] we can use the recognition and generative densities to define a single density $R$ over configurations of the system, factored into "wake" and "sleep" cycles:

$$R(s_i = 1) = S(\text{wake})Q(s_i = 1) + (1 - S(\text{wake}))P(s_i = 1) \quad (5)$$

Using this model, the identity thesis can be formulated as the claim that the VFE is equal to $FE_R$, the (thermodynamic) Helmholtz free energy of the system when the probabilities of the configurations are as given by $R$ of Eq. (6), which specifies the (marginal) probabilities $R_i$ of each unit $s_i$'s firing, regardless of direction of influence. This equivalence is easily seen to hold when the VFE is minimized, since the recognition and generative densities will in that case be equal and probabilities in the network will conform to a Boltzmann distribution.

With respect to non-equilibrium states, we may still expect $FE_R$ to scale with the VFE: to the extent that there is nonzero K-L divergence between $P$ and $Q$, the bottom-up and top-down drives will favor different configurations of the system, and the entropy of $R$ and thus the surprisal of the internal states will be greater. Moreover, lowering the entropy of $R$ (and thus the VFE) as $FE_R$ is simultaneously minimized entails lowering the internal energy, assuming a constant temperature.[11]

---

[10] Here the form of the distribution $S$ is not important. In practice, $S(\text{sleep}) = S(\text{wake}) = 0.5$.

## 2.2 The free energy identity thesis

The preceding argument is useful as a first approximation, but the phases of the wake-sleep algorithm are biologically unmotivated [39], and, perhaps more importantly, the algorithm is doubly an approximation to Bayesian inference in that the update procedure for the recognition model does not exactly minimize the free energy used to define the cost ([50], p.17).[12]

As discussed in the latter reference, the wake-sleep algorithm can be viewed as an approximation to the E-M algorithm, which in [3] is given an interpretation in terms of VFE minimization.[13] The predictive coding model of [5] offers an exact implementation of E-M in which the "expectation" step corresponds to approximate inference of hidden causes from sensory input and the "maximization" step adjusts (reciprocal) synaptic weights to fit both generative and recognition densities to the estimated states, and thereby to the sensory input. In this model, the top-down influences of *state* or representation units at one level on *error* units in the level below (mediated by top-down synaptic weights) "instantiate the forward model" (p.823), while the bottom-up connections from error units to higher-level state units implement a "recognition" term, which doubles as the likelihood under the generative density. Minimizing the VFE is equivalent to minimizing the prediction error (p.821).

As noted above, this representation entails that the VFE depends only on the current state of the system. The reasoning of the previous section could, however, be extended to a model that has the virtues of predictive coding (as described in [5]) over wake-sleep (e.g. tied forward and backward weights and, if desired, lateral connections and explicitly represented prediction errors) while using a stochastic encoding of the variances, so that the representation of the VFE is distributed over alternative configurations of the system and it is thus converted into an ensemble property.

---

[11] Of course, some of the internal energy will be lost as heat and excluded from the Helmholtz free energy. The argument given here assumes that this "bound" portion of the energy is also the representationally "meaningless" portion. For example, it ought to be the case that less of the total energy input to a processing layer is squandered, to the extent that units with stochastic sigmoid activation functions are further from their saturation points.

[12] One problem lies in the "sleep" phase updates, which in fact minimize the K-L divergence between *P* and *Q*, rather than the divergence between *Q* and *P* (not necessarily equal) that appears in the free energy. Wake-sleep is fairly effective despite this and other approximations, but approximate VFE minimization obviously will not do for our purposes.

[13] E-M may be viewed as an alternative to full variational Bayes in which a point estimate is computed for the parameter values.

The arguments given above may be bolstered by examining an alternative way of writing the variational lower bound (negative free energy) $F$, considered in connection with varational autoencoders [2]. A similar analysis occurs in the literature on the FEP and active inference [46], where it appears as the relation *free energy = complexity – accuracy* [51].[14] In terms of $-F$ or free energy:

$$-F = D_{KL}(Q(h|v)||P(h)) - \sum_{h} Q(h|v) \log P(v|h) \qquad (6)$$

Maximizing the sum in the second right-hand term in (6) ensures accurate reconstruction of the inputs, and thus minimizes the prediction error incurred when hidden states are inferred from observations and then used for prediction. Minimizing the first term ensures that the network will, absent external prompting, tend to occupy the same states that are sampled from the recognition density when the network is driven by sensory input. Together, these terms maximize the expected similarity between states determined by the network's internal dynamics and those induced by sensory input, minimizing perturbations caused by the inputs.

Parr et al [18] show that under certain assumptions, changes in free energy are approximately equal to changes in the joint surprisal of the states inside the system's defining Markov blanket as well as the "active" and "sensory" states comprising the blanket.[15] These quantities also track the amount of heat dissipated due to a corresponding change in free energy.[16] The approximation ignores the entropy over the recognition density, which (under the assumption that the density is Gaussian near its mode) can be expected not to change much near non-equilibrium steady state, but otherwise this result agrees with that of the reasoning just rehearsed.

Since the identity thesis requires a more far-ranging equivalence, we might attempt to bring the recognition entropy into the picture as follows. An overarching principle linking thermodynamics to variational inference, implicit in the arguments above, is that the degree of Bayesian updating required to maximize the inferential coherence among beliefs will (given our modeling assumptions) be reflected in proportional expenditures of energy in the physical sense. In Friston's words, "statistical and thermodynamic efficiency go hand-in-hand" ([9], p.120). Learning will reduce the average size of belief updates by minimizing the divergence between generative and recognition densities, but it will also *maximize* the recognition entropy, subject to the constraint that the two densities converge. This provides an independent way of reducing the size of the updates, and so the thermodynamic energy expenditure: holding the model evidence constant, a "softer", higher-entropy stochastic code requires less physical energy per code to represent input $v$ using states of hidden units $h$ than a "hard" code in which $p(h|v)$ tends to be closer to 0 or 1, requiring stronger excitatory or inhibitory signaling.

---

[14] Action can modify the accuracy by selectively sampling states with high likelihood under the generative model. Thus, not surprisingly, the identity thesis proposed here extends implicitly to action as well as perception and cognition. Thanks to Jakob Hohwy for drawing my attention to this connection.

[15] For present purposes, the "active" and "internal" states in the sense relevant to Markov blankets can be grouped together as "hidden" states.

[16] [18] includes a precise analysis of the relationship between heat and surprisal or self-information, which trades on the fact that spontaneous processes that involve increases in entropy are irreversible, and thus a system will not spontaneously follow the reverse path in its phase space. "Roughly, the amount of heat dissipated…along a given path is an expression of how surprising it would be to observe a system following the same path backwards relative to forwards" (p.9).

## 3 The identity thesis: discussion

*3.1 From isomorphism to identity*

One way of attempting to express the content of the identity thesis is with the formula $VFE = A$, or in terms of dynamics, $\Delta VFE = \Delta A$, where $A$ is the thermodynamic Helmholtz free energy. However, the arguments given above license only a mathematical interpretation of these equalities, not a logical interpretation in terms of strict identity. Securing the identity thesis proper requires more philosophical heavy machinery.

Smart's argument for the identity thesis [14] was essentially an appeal to Ockham's razor: assuming we have evidence for compelling correlations between mental states and brain states, we should assume them to be identical barring specific reason not to do so. The bulk of Smart's paper is dedicated to convincing the reader that there are no such defeating reasons.

This kind of argument can be made more rigorous by appeal to David Lewis's brand of functionalism in the philosophy of mind [15]. Lewis begins by considering a hypothetical collection of "platitudes" or truisms about mental states such as beliefs, desires, and sensations—generalizations about which combinations of such states are likely to lead to which others, which states are likely to follow upon certain sensory stimuli and eventuate in certain kinds of behavior, and so on, as well as statements about which state types subsume other types, such as "seeing red is an instance of perceiving". Taken together, these truisms plausibly capture the content of the intuitive or "folk" psychological theory that we bring to bear when we explain and predict human behavior in terms of inner states in everyday contexts.

Beginning with the logical conjunction[17] of the "platitudes", we may replace the mental-state terms by variables $x, y, z$, etc. and prepend appropriate existential quantifiers, obtaining a lengthy statement to the effect that a certain collection of things, related to one another in the ways specified by the platitudes, exists—technically, a formula in first-order predicate logic (a Ramsey sentence) of the form $(\exists \mathbf{x})T(\mathbf{x})$.[18] This sentence captures the relations among the inner states described by the folk theory without explicitly referring to them (see [15], p.253).

---

[17] What Lewis suggests is in fact not taking the conjunction of *all* folk-psychological platitudes as our implicit definition but rather "the conjunction of most disjunctions of most of them", so that a few false platitudes will not falsify the identity.

Since commonsense psychology posits definite collections of inner states ("Smith's beliefs", i.e. "*the* beliefs of Smith"), rather than simply stating that at least one such collection of states exists, we can supplement the Ramsey sentence above with a (Russellian) uniqueness condition, yielding a *modified Ramsey sentence* declaring that the roles specified by the predicate $T$ are uniquely realized by the values of $\mathbf{x}$:

$$(\exists \mathbf{x})(\forall \mathbf{y})[T(\mathbf{y}) \equiv \mathbf{x} = \mathbf{y}]$$

Here, '≡' denotes the material biconditional and '=' denotes the identity relation.

This sentence implicitly defines the mental states in terms of the causal roles they occupy with respect to one another (and with respect to observable behavior and stimuli).[19] If mature neuroscience discerns brain states that in fact occupy these causal roles, Lewis argues, we have found the mental states, since in order to be a mental state it is (by definition) sufficient for something to play the causal/functional role described in the Ramsey sentence. The identity of psychologically and neuroscientifically described states then follows as a matter of logic.

We are now in almost precisely the position Lewis envisioned, but in place of folk-psychological platitudes, we have a sophisticated, quantitatively expressed cognitive theory (i.e. the description of a system in terms of variational Bayesian inference over a collection of hypotheses), in which the equations governing hypothesis updates and the evolution of their conditional probabilities specify the relevant causal roles. This theory yields the left-hand side of the equation $\Delta VFE = \Delta A$. On the right-hand side, we appeal not so much to neuroscience *per se* as to the underlying physics, which we can relate sensibly to brain states thanks to theoretical advances in neural network modeling, the theory of self-organization and control theory [44], and non-equilibrium steady state physics [53].

The normative (Bayesian) as opposed to descriptive character of the cognitive theory does not signal as deep a difference from Lewis's original program as it may initially appear to—the relevant assumption is that systems of the right sort will predictably exhibit internal processes isomorphic to (approximate) Bayesian computations, not that they ought to do so.[20] Nor is it suspicious that the cognitive theory in question was originally derived in part via an abstraction from neuroscience itself. Indeed, if the mind *just is* the brain (and vice-versa!), this convergence between cognitive and physical descriptions, once both are maximally informed, is precisely what one would expect.

---

[18] Here, $\mathbf{x}$ is a vector of bound variables $x_1$, $x_2$, etc., and $T$ is a predicate expression semantically equivalent to the conjunction of the open sentences corresponding to the individual platitudes.

[19] If the content of a representation (for instance, seeing *red* as opposed to seeing in general) is determined by its functional role (as argued in the context of generative models in [52]), then content-individuation of mental states is seen to be a fine-grained way of functionally individuating states, and thus falls out of Lewis's approach to psychophysical identification as well.

In assuming that mental states are the internal states responsible for causing action, and *a fortiori* that they are internal states of organisms, we beg the question against a radical mind-body dualism, according to which mental states lack spatiotemporal locations. What is exciting about an identity thesis in 2020, however, is not so much its opposition to Cartesian dualism (many these days are committed functionalists or identity theorists in any case), but (a) the fact that the connection between mind and body is drawn at the level of *physics*, undercutting even neuroscience as a reduction base, (b) that the reasons for which, and the precise way in which, the identity is realized can be discerned, at least in broad strokes, on the basis of current science, and (c) that the theory is, accordingly, expressed in quantitative terms.

*3.2 Scope of the thesis*

The free energy principle applies very broadly to any physical system that can be understood as possessing a Markov blanket, including brains but also simple self-regulating systems such as cells, larger systems such as animal populations and entire ecosystems, and presumably most everything in between, including, perhaps, human social institutions [11,13]. Although the arguments above focused on the case of the brain, they include in their scope any system in which a generative model can be regarded as stochastically encoded in the dynamics of the internal elements. This seems to imply something close to panpsychism[21] if one takes the sort of statistical inference in question to be sufficient for mentality (see [54] for discussion).

Terms of art such as "Bayesian belief" are sometimes used to frame hypotheses in cognitive science without commitment to such radical conclusions (cf. the subpersonal representations [55], low-grade forms of intentionality [56], and mere computational states of previous eras). One may wish to restrict the extension of the term "mind" to those systems that possess generative models of some specified degree of complexity, capable of supporting certain special classes of beliefs [13,18,19,30,57].[22] It is not clear, however, that this strategy provides more than verbal comfort: if the "Bayesian beliefs" encoded in the stochastic thermodynamics of the human brain provide a compelling explanation of human cognition and behavior, they do so simply in virtue of their status as subjective probability distributions over external states, a status shared by the "beliefs" of protozoa. However this argument goes, mind-brain identity will fall out as an entailment of the (potentially broader) V*FE* / *A* identity thesis.

---

[20] The rationality of Bayesian inference does of course furnish part of the argument as to *why* we should expect adaptively successful systems to exhibit this internal structure, but is no part of the structure itself.

[21] In the sense that psychological properties would be universal or very widespread, in contrast to a contemporary usage of this term to refer specifically to a thesis about consciousness.

[22] See [58] for a more deflationary take on minds.

Deeper than the joint in nature between complex, human-like agents and simpler ones, however, is that between systems whose internal models are implemented *transparently* by local interactions among their physical parts (where such interactions have typically been sculpted by self-organization in response to external pressures—see [59]), and those in which this is not the case. Ensembles of neurons in biological systems provide one example of the former sort of self-organization. Simulated neural networks may exhibit the requisite computational dynamics, which can be expected to some extent to be reflected in their thermodynamics (as exhibited in the overheating of a laptop used to train a deep learning model), but to the extent that software-hardware relations in digital computers involve arbitrary conventions or inefficient use of the physical resources, there will be a mismatch between physical and variational energetic descriptions that voids application of the identity thesis.

As an example, compare the digital simulation of neural networks to neuromorphic analog computation.[23] To consider just one particularly salient difference, the (potential) energy required to store the state (e.g. instantaneous firing rate) of a single neuron or subpopulation, usually represented numerically in a simulation as a real value greater than zero, will in many digital representations be constant for all values, within some allowable range determined in advance by various modeling and implementation choices. Under the stochastic encoding of probabilities discussed in §2, the same amount of energy will thus be required to represent any positive probability that figures in the variational free energy. In real neuronal networks or physically analogous computers, by contrast, the energy used so to encode a probability scales with the probability.[24] The identity thesis may then provide a criterion for distinguishing minds from relatively crude simulations of them, ruling out many systems whose mentality would offend intuition most, like current software implementations of autoencoders, or corporations—though not prokaryotes or galaxies.

An intimately related route to transparency appeals to the scale-free nature of variational free energy dynamics. As suggested in [11,12,13], the relations between global and local free energy minimization in a supersystem can be understood in terms of subsystems themselves undergoing the same type of energy minimization/inference process at their own scales, yielding hierarchies of embedded systems all of which obey the FEP, and in which the dynamics of superordinate systems constrain those of the subsystems [13][25] (see [62] for an application to neuronal ensembles). Since the thermodynamic potential energies of a complex system and of its subsystems are constitutively related, the identity thesis naturally yields such mutual constraints on inferential dynamics across levels of organization.

---

[23] Many thanks to Thomas Parr for suggesting this example.

[24] Of course, under minimal assumptions, any physical system including a digital computer has a Markov blanket, and thus an associated "extrinsic" description in terms of variational inference [9,19]. The relevant point is that this interpretation of the physical system (the hardware) needn't match the cognitive description conventionally associated with the software level (i.e. the simulated network).

**Conclusion**

In this paper, I've canvassed the prospects for the hypothesis that brains and other living systems implement variational inference as a necessary consequence of potential energy minimization in the thermodynamic sense. A close parallel from the history of science is the identification of inertial and gravitational mass, whose establishment on empirical grounds long antedated its theoretical explanation, thanks to the *a priori* distinctness of the relevant concepts.[26] In contrast to that case, I have given no axiomatic proof of the equivalence of thermodynamic and informational free energy, but have attempted rather to describe conditions under which such an identification is plausible.

The more formal treatments in [9] and [18] demonstrate an approximate identity or constraint between thermodynamic potential energy and VFE near non-equilibrium steady state. [19] defends a position similar to the view sketched in this paper, but with differences of detail and interpretation. Most significantly perhaps, the isomorphism or identity advanced in the latter concerns the "intrinsic" and "extrinsic" information geometries of the internal states of a system, which together specify how changes in internal states over time map to changes in belief. The present argument is meant to support a slightly less abstract isomorphism or identity, directly between variational and thermodynamic free-energy descriptions. This argument is consistent with the detailed treatment of the metabolic efficiency of variational inference given in [45].

A key test for the identity thesis is whether a stochastic code of the kind discussed in §2.1 turns out to be empirically plausible. This constraint on a choice of encoding for statistical models demonstrates one respect in which the identity thesis is theoretically useful: the conjunction of the latter with the FEP is a falsifiable hypothesis, even if the FEP itself is not [63]. Of course, the argument may run in the other direction: strong empirical evidence for the right type of encoding, together with a formulation of cognitive dynamics in terms of variational Bayesian inference, supports the case for "variational psychophysical identity".

---

[25] This perspective recalls the "homuncular functionalism" defended in philosophy by William Lycan [60], as well as Leibniz's remark that "the organic body of each living being is a kind of divine machine or natural automaton, which infinitely surpasses all artificial automata. For a machine made by the skill of man is not a machine in each of its parts" ([61], §64).

[26] Thanks very much to Maxwell Ramstead for suggesting this comparison.

**Acknowledgements**

Thank you to Jakob Hohwy, Maxwell Ramstead, and Thomas Parr for discussion and for early comments on a draft of this paper, to Karl Friston for helpful correspondence, to two anonymous reviewers for their constructive suggestions, and to Mel Andrews and Michael Kirchhoff for discussions of related work.